\documentclass[prl,reprint,twocolumn,preprintnumbers,amsmath,amssymb,showpacs,footinbib,superscriptaddress]{revtex4-1}

\usepackage[titletoc,toc,title]{appendix}
\usepackage{braket}
\usepackage[final]{feynmp}
\usepackage{ifpdf}
\usepackage{comment}
\usepackage{amsmath}
\usepackage{mathrsfs}
\usepackage{color}
\usepackage{bm}
\usepackage[font=small]{subcaption}
\usepackage[font=small]{caption}

\graphicspath{{./Figures/}}

\makeatletter
\def\endfmffile{%
	\fmfcmd{\p@rcent\space the end.^^J%
			end.^^J%
			endinput;}%
	\if@fmfio
		\immediate\closeout\@outfmf
	\fi
	\ifnum\pdfshellescape=\@ne
		\immediate\write18{mpost \thefmffile}%
	\fi}
\makeatother

\usepackage[demo]{graphicx}
\usepackage{dcolumn}
\usepackage{bm}
\usepackage{hyperref}

\newcommand{\beq}{\begin{equation}}
\newcommand{\eeq}{\end{equation}}

\renewcommand{\thanks}[1]{\footnote{#1}} 

\newcommand{\be}{\begin{equation}}
\newcommand{\bea}{\begin{eqnarray}}
\newcommand{\eea}{\end{eqnarray}}

\newcommand{\ee}{\end{equation}}

\def\ba{\begin{eqnarray}}
\def\ea{\end{eqnarray}}

\def\14{{1\over4}}
\def\12{{1 \over 2}}

\def\8{\infty}
\def\oh{\frac{1}{2}}

\def\undertext#1{\vtop{\hbox{#1}\kern 1pt \hrule}}

\def\VEV#1{\left\langle\,#1\,\right\rangle}

\def\be{\begin{equation}}
\def\ee{\end{equation}}
\def\bea{\begin{eqnarray} & &}
\def\eea{\end{eqnarray}}

\def\rf#1{(\ref{#1})}

\def\rf#1{(\ref{#1})}

\begin{document}


\title{Randomly measured quantum particle}

\author{Victor Gurarie}
\affiliation{Department of Physics and Center for Theory of Quantum Matter, University of Colorado, Boulder, Colorado 80309, USA}

\begin{abstract}
We consider the motion of a quantum particle whose position is measured in random places at random moments in time. We show that a freely moving particle measured in this way undergoes superdiffusion, while a charged particle moving in a magnetic field confined to the lowest Landau level undergoes conventional diffusion. We also look at a particle moving in one dimensional space in a random time-independent potential, so that
it is Anderson localized, which is also  measured at random points in space and randomly in time. We find that random measurements break localization and this particle also undergoes diffusion. To address these questions, we develop formalism similar to that employed when studying classical and quantum problems with time-dependent noise. 
\end{abstract}

\maketitle



It became widely appreciated in recent years that dynamic evolution of a quantum system, when it is subject to  measurements, acquires new features
 \cite{Fisher2018, Nahum2019}. Under the projective measurements the state of a quantum system changes according to
\be \left| \psi_i \right> \rightarrow \left| \psi_f \right> = \hat P_n \left| \psi \right>.
\ee Here $\hat P_n$ is a projective measurement operator obeying $\hat P_n^2=\hat P_n$, and $n$ is the outcome of the measurement, such that
\be \sum_n \hat P_n = 1.
\ee It is common not to normalize the wave function post measurement, so that the probability of a particular measurement outcome is given by $p_n=\left< \psi_f \right| \left. \psi_f \right>$. 

A situation that we may have in mind involves a system governed by a Hamiltonian $\hat H$ evolving for some time  which is then followed by a projective measurement. After that, this procedure is repeated $N$ times. The final unnormalized wave function of the system is then given by
\be \label{eq:psif} \left| \psi_f \right> = \hat P_{n_N} e^{-i \hat H t_N} \hat P_{n_{N-1}} e^{-i \hat H t_{N-1}} \dots \hat P_{n_1} e^{-i \hat H t_1} \left| \psi_i \right>.
\ee
The probability of the particular  outcome of this series of measurements labeled by $n_N, n_{N-1}, \dots, n_1$ is given by
$ \left< \psi_f \right| \left. \psi_f \right>$, 
so that average of an observable  $\hat Q$ is given by
\be \label{eq:q} Q = \sum_{n_N, n_{N-1}, \dots, n_1} \left< \psi_f \right| \hat Q \left| \psi_f \right>.
\ee
Note that unless $\hat Q$ commutes with both the measurements $\hat P_j$  and the Hamiltonian $\hat H$, averaging over the outcome of the measurements is not equivalent to doing no measurements at all. 

The procedure described here is straightforward but not easy to work with in analytic calculations. Instead of working with the projective measurements, let us work with ``weak measurement" operators, defined in such a way that they deviate just slightly 
from the identity operator \cite{Murciano2023,Garratt2023,Poboiko2023,Guo2024}. A general definition of such operator may consists of entangling the system with an ``ancilla" system whose eigenstates are labelled by an integer $n$. A unitary evolution may evolve the system from a state $ \left| \Psi_i \right> = \left| \psi_i \right>\otimes \left| 0 \right>_a$ to the state $\left| \Psi_f \right> = \sum_n \left[\hat K_n \left| \psi \right> \right]  \otimes \left| n \right>_a$. Here the state of the ancilla is labelled by a subscript ``$a$", and the initial state of the ancilla is denoted $\left| 0 \right>_a$. Measuring the ancilla leaves it in a state  $n$, which results in the system now being in the state 
$\left| \psi_f \right> = \hat K_n \left| \psi \right>$. In order for a unitary operator to exist which can
evolve $\left| \Psi_i \right>$  into $\left| \Psi_f \right>$ for all $\left| \psi_i \right>$, we need $ \left| \Psi_f \right>$ to have unit norm, which immediately leads to
\be \label{eq:kraus} \sum_n \hat K_n^\dagger \hat K_n = 1.
\ee
Such operators $\hat K_n$ represent the examples of Kraus operators. In the context discussed here, they generalize the projective measurement operators $\hat P_n$. But now, unlike $\hat P_n$, $\hat K_n$ could be close to the identity 
operator.

 As a concrete example, suppose we would like to study a particle of mass $m$ moving in a potential $U(x)$, subject to measurements to determine if this  particle is located in a particular region in space. Instead of projecting onto this region, consider the operator
\be \hat K[V]  = \int dx \, e^{V(x) \delta t} \left| x \right> \left< x \right|.
\ee The time interval $\delta t$ is inserted here for convenience. 
The larger the value of $V(x)$ nearby a particular point $x$, the more likely it is to find the particle nearby this position after $\hat K$ is applied to a state of the particle. Different functions $V(x)$ correspond to the different
measurement outcomes, so they generalize the index $n$ used to label different measurement outcomes above. Just like in Eq.~\rf{eq:psif}, we will let the system evolve, then apply $\hat K$  with a particular $V(x)$, then repeat the process with a new function $V(x)$. 
This can be encoded by functions $V_j(x)$ where $j$ labels the instances in time when the measurements occur. 

Summation over $n$ can now be promoted to integration over the functions $V_j(x)$ with a suitable weight. If measurements occur at random times, we can take $V_j(x)$ as random and Gaussian in both time and space with the correlation function
\be \label{eq:gaus}  \VEV{V_{j_1}(x_1) \, V_{j_2}(x_2) } =\frac{\lambda}{\delta t} \, \delta_{j_1 j_2} \, W\left(x_1-x_2 \right).
\ee 
Here $\lambda$ is the ``strength" of the measurements, while the function $W$ goes to zero as its argument becomes much larger than 1. We set $W(x)=g(x/\ell)/\ell$, with the function $g$ satisfying
\be \label{eq:w} \int dx \, g(x) = 1.
\ee
This effectively fixes the length of the interval where the measurement takes place to be $\ell$. In the limit $\ell \rightarrow 0$ we could replace $W$ by the delta function. In what follows we will often take the limit $\delta t \rightarrow 0$, which promotes  $V_j(x)$ to a function of both space and time $V(x,t)$ with the correlation function
\be \label{eq:gaus1}  \VEV{V(x_1,t_1) \, V(x_2,t_2) } =\lambda \, \delta (t_1-t_2) \, W\left(x_1-x_2 \right).
\ee

We can construct the functional integral describing the wave function of this particle according to the standard Feynman first-quantized path integral
\be \label{eq:feyn} \psi_f(x_f, t_f) = \int^{x(t_f)=x_f}_{x(t_i)=x_i} {\cal D} x(t) \, e^{i S[x]+ \int dt \, V(x, t) }.
\ee Here \be S[x] = \int dt \left( \frac{m \dot x^2}{2} - U(x) \right) \ee is the action of the particle.

Defined in this way, together with random $V$, the operators $\hat K$ defined above when averaged over random $V$ satisfy the Kraus operator condition \rf{eq:kraus}. Indeed, working
with the discrete time again, we have
\be \hat K^\dagger \hat K = \int dx \, e^{2 V(x) \, \delta t} \left| x \right> \left< x \right|. 
\ee Summation over $n$ in \rf{eq:kraus} is equivalent to averaging over random $V$ using the correlation function \rf{eq:gaus}, with the result
\be \label{eq:kk} \VEV{\hat K^\dagger \hat K} =  e^{ 2 \lambda W(0) \delta t} \hat I.
\ee
where $I$ is identity operator. The proportionality constant in front of it signifies that the probabilities (density matrices) produced by \rf{eq:feyn} will still need to be normalized before expectations of any operators 
can be computed.

The formalism that we now constructed is equivalent to quantum mechanics of a particle moving in a purely imaginary potential random in both time and space.  The potential being imaginary is novel aspect of this problem, reflecting the non-unitary nature of measurements. But motion 
in a potential random in both space and time is a well known problem, investigated in the past for example in the context of the directed polymer problem and KPZ equation \cite{Kardar1987,Kardar1987a,Calabrese2010}. 

We would like to construct a functional integral which would produce a density matrix $\rho(x_+, x_-)= \psi_f(x_+) \psi_f^*(x_-)$ which could then be averaged over random $V$ and used to compute averages of observables such as in \rf{eq:q}. We find
\be \label{eq:ke}  \rho = \int {\cal D} x_+ {\cal D} x_- e^{i S[x_+] -i S [x_-] + \lambda \int dt \, W \left(x_+(t) - x_-(t) \right)}.
\ee
We arrive at a picture of a Keldysh-like functional integral \cite{Kamenev2011}, with interacting forward and backward evolving fields. We can now rely on this framework to solve several problems. For applications, we recast this functional integral 
in the form of a Schr\"odinger equation for the density matrix in the Choi-Jamio\l kowski  representation \cite{Jiang2013}
\begin{eqnarray}  \label{eq:schr} i \dot \rho(x_+, x_-) &=&  \hat H_+ \rho(x_+,x_-) - \hat H^*_- \rho(x_+,x_-) + \cr && i \lambda W(x_+-x_-) \rho(x_+, x_-).
\end{eqnarray}
Here $\hat H^*$ represents the complex conjugate Hamiltonian.

As a first application, consider a freely moving particle with the Hamiltonian $\hat H = \hat p^2/(2m)$ which is measured at random times and places. Its density matrix satisfies the Schr\"odinger equation
\be \label{eq:schr2}  i \dot \rho =  - \frac{1}{2m} \frac{\partial^2 \rho}{\partial x_+^2} + \frac{1}{2m} \frac{\partial^2 \rho}{\partial x_-^2}  + i \lambda W(x_+-x_-) \rho.
\ee
Suppose initially the particle is confined to  a particular region in space, $\psi_i (x) =\exp\left(-x^2/(4 \Delta^2) \right)/\left(2 \pi \Delta^2 \right)^{1/4}$. With this initial wave function, $\left< \psi_i \right| \hat x^2 \left| \psi_i \right>=\Delta^2$. 
In other words, we need to solve \rf{eq:schr2} with the initial conditions
\be \rho_i = \frac{1}{\sqrt{2 \pi} \Delta}  \exp\left( - \frac {x_+^2+x_-^2}{4 \Delta^2} \right).
\ee
This problem is exactly solvable for any function $W$. To solve it, we introduce the convenient variables $x_{cl} = (x_++x_-)/2$, $x_{q} = x_+-x_-$, to rewrite the equation as
\be i \dot \rho = \left(- \frac 1 m \frac{\partial^2}{\partial x_{cl} \partial x_{q}} + i \lambda W(x_q) \right) \rho.
\ee
We look for the eigenfunctions and eigenvalues of the operator on the right hand side. Since it is not Hermitian, we need to work out both right and left eigenstates of this operator, satisfying
\be \left(- \frac 1 m \frac{\partial^2}{\partial x_{cl} \partial x_{q}} + i \lambda W(x_q) \right) \psi_R = E\psi_R,
\ee
for the right eigenstate. 
These can be found explicitly
\be \psi_R(q,k) = \frac{1}{\sqrt{L_q L_{cl}}} e^{i q x_{cl} + i k x_q -\frac{\lambda m x_q}{q L_q} + \frac{\lambda m s(x_q)}{q}},
\ee
where $s'(x) = W(x)$, $s(x)=-s(-x)$. $L_q$ and $L_{cl}$ are the range of the variables $x_q$ and $x_{cl}$ respectively, and  clearly (thanks to \rf{eq:w}) $s(\pm L_q)=1/2$. 
The left eigenfunction is given by
\be\psi_L(q,k) = \frac{1}{\sqrt{L_q L_{cl}}} e^{-i q x_{cl} - i k x_q +\frac{\lambda m x_q}{q L_q} - \frac{\lambda m s(x_q)}{q}},
\ee
The corresponding eigenvalues are
\be E= \frac{q k}{m} + i \frac{\lambda}{L_q}.
\ee
The time evolved $\rho$ can be found by projecting the initial state $\rho_i$ onto the eigenfunctions and evolving those in time using the eigenvalues $E$. After some algebra we arrive at the following expression
\begin{align} &   \rho(y_{cl} , y_q, t) =  \label{eq:rho} \\ &  \int \frac{dq}{2\pi} e^{{i q y_{cl} + \frac{\lambda m}{q} \left( s(y_q)-s(y_q-qt/m) \right) - \frac{4 q^2 \Delta^4 + \left(y_q - qt/m\right)^2}{8 \Delta^2}} } .
 \nonumber \end{align}
This could be used to evaluate  expectations of any operators at the time $t$. In particular, let us evaluate the average square of the position of the particle
\be \VEV{y^2} = \int dy_{cl} \, y_{cl}^2 \, \rho(y_{cl},0,t) / \int dy_{cl} \rho(y_{cl},0,t).
\ee
The division by the normalization factor is needed to compensate for the lack of normalization of the $\hat K$ operators due to the right hand side of \rf{eq:kk}. 

It is straightforward to calculate this from \rf{eq:rho}. We need to define two coefficients $\alpha$ and $\beta$ from the expansion
\be s(x) \approx \alpha x/\ell-\beta x^3/\ell^3 + \dots.
\ee
The answer is then expressed in terms of these as
\be \label{eq:res} \VEV{y^2} = \Delta^2 + \frac{\hbar^2 t^2}{4 m^2 \Delta^2} + \frac{2  \beta \lambda t^3}{m^2 \ell^3}.
\ee
This constitutes the answer to the problem of randomly measured particle on a one dimensional line. Here to further elucidate the meaning of this answer we restored the Planck constant $\hbar$ in it. 

In this answer the first term represents the initial uncertainly in the position of the particle. The second term is the spread of the initial wave packet due to the usual quantum effects. The last term is new and is due to the particle being repeatedly measured at random times in random positions, on intervals of the length $\ell$. Note that this term is purely classical.

Clearly the last term describes superdiffusion, as the displacement $R \sim t^{3/2}$. It is straightforward to verify that this is the correct physics of a randomly measured particle. Every time the particle is measured, its velocity acquires a random boost of the typical magnitude of $\hbar/(m \ell)$. The velocity 
therefore undergoes Brownian motion, which implies that the position of the particle after $n$ measurement events is given by
\be R \sim \sum_{j=1}^N \delta t \sum_{i=1}^j \frac{\hbar n_i}{m \ell},
\ee where $n_i=\pm 1$ are random variables determining the direction of the velocity boosts. 
We see from here that the average square of the position after $n$ steps is given by
\be R^2 \sim \frac{\hbar^2 \delta t^2}{m^2 \ell^2} n^3 = \frac{\hbar^2 \delta t^2}{m^2 \ell^2} \left( \frac{t}{\delta t} \right)^3. 
\ee
We can now estimate $\lambda \sim \ell \hbar^2/ \delta t$, which gives
\be R^2 \sim \frac{ \lambda t^3}{m^2 \ell^3}.
\ee
This matches the exact result above \rf{eq:res}.

To further illustrate the power of the approach via the equation \rf{eq:schr}, let us now consider a particle moving in two dimensions in a uniform magnetic field, which is randomly measured while remaining confined to the
lowest Landau level. This implies that the measurements determine if a particle is located within randomly located domains of the area no less than the square of the magnetic length. 

We choose the Hamiltonian in the symmetric gauge,
\be \hat H = - \frac{1}{2M} \left[ \left(\partial_x + \frac{i e}{2 c} B y \right)^2+ \left( \partial_y- \frac{i e}{2  c} B x\right)^2 \right].
\ee
As always in this problem it is convenient to introduce the dimensionless coordinates $X=x/\sqrt{2} \ell$, $Y=y/\sqrt{2} \ell$, where $\ell = \sqrt{ c/eB}$ is the magnetic length. 
It is also common to introduce the complex coordinates $z=X+iY$. 
In terms of these
the lowest Landau level wave functions are given by \cite{PrangeGirvinQHE}
\be \psi_n = \sqrt{\frac{1}{ \pi n!} } \, z^n e^{- \oh z \bar z}.
\ee 
We expand the density matrix in the basis of these eigenfunctions
\be \rho = \sum_{mn} c_{mn} \psi_m \psi^*_n,
\ee
and project the equation \rf{eq:schr} onto this basis. It is convenient to take $W = \delta(x)\delta(y)$, as this simplifies algebra while projection to the lowest Landau level guarantees that the measurements are done on the scale of the magnetic length $\ell$. 

Let us take $\rho_i = \psi_0 \psi_0^*$ as the initial density matrix. This is invariant under rotations, while any product $\psi_n \psi^*_m$ with $n \not = m$ is not. Therefore, $\rho$ remains diagonal during its time evolution and we can write
\be \rho = \sum_n c_n \psi_n \psi^*_n.
\ee

We find
\be  \dot c_n =\lambda  \sum_m c_m \int d^2 z \, \psi_n \psi_n^* \psi_m \psi_m^*.
\ee
Evaluating the integral produces the equation
\be  \dot c_n = \frac{\lambda}{2\pi} \sum_m \frac{(n+m)!}{2^{n+m} n! m! } c_m.
\ee
For the initial conditions we take $c_n = \delta_{n0}$. In other words, the particle initially is in the state $\psi_0$, which represents the state with the smallest possible spatial extend in the lowest Landau level. We note that
\be  \sum_{n=0}^\infty \dot c_n =\frac{\lambda}{2\pi}   \sum_{n=0}^\infty \frac{\lambda}{2\pi} \sum_m \frac{(n+m)!}{2^{n+m} n! m! } c_m = \frac{\lambda}{\pi} \sum_{m=0}^\infty c_m.
\ee
The solution of this equation
\be \label{eq:no}  \sum_{m=0}^\infty c_m = e^{\frac{\lambda t}{\pi}}
\ee represents the overall normalization factor which will need to be divided by at the end of the calculation. 

Now noting that
\be \left< n \right| \hat r^2 \left| n \right> = n+1 \nonumber
\ee 
we define $d=\sum_{n=0}^\infty (n+1) c_n$ and calculate $\dot d$ to find
\be \sum_{n=0}^\infty (n+1) \sum_{m=0}^\infty \frac{(n+m)!}{2^{n+m} n! m! } c_m = \frac{\lambda}{\pi} \sum_{m=0}^\infty (m+2) c_m. \nonumber
\ee
This gives
$ \dot d = (\lambda/{\pi }) \left( d+ e^{\frac{\lambda t}{\pi}} \right)$, with the solution
$ d = e^{\lambda t/\pi} \left( 1  + {\lambda t}/{\pi} \right)$. 

Restoring the magnetic length and dividing by the overall normalization  \rf{eq:no} gives for the average square of the displacement
\be R^2 = 2\ell^2 \left( 1  + \frac{\lambda t}{2\pi \ell^2} \right).
\ee
This is clearly represents diffusion. 

Finally, let us examine a problem of a particle moving in a random time independent potential whose position is randomly measured \cite{Gefen2023}. Let us stay in one dimensional space, where random time independent
potential will Anderson-localize all eigenstates. In the absence of random measurements a particle placed in a particular point will remain localized and will not move far from this point. If the particle is
randomly measured, it can start hopping from one localized eigenstate to another. This can be captured with the 
same Schr\"odinger equation \rf{eq:schr} with the Hamiltonian including a random time independent potential. Such an equation is difficult to solve analytically. We discretize it in space and solve it numerically. 
The equation we solve is
\begin{eqnarray} && i \dot \rho_{mn} =J \left( \rho_{m,n+1}+\rho_{m,n-1} - \rho_{m+1,n} - \rho_{m-1,n} \right)+ \cr && \left(U_m - U_n \right) \rho_{mn} +i \lambda \delta_{mn} \rho_{mn},
\end{eqnarray}
where $J$, the hopping matrix elements, are set to 1, $U_n \in [-3,3]$ are random time independent variables large enough to ensure tight eigenstate localization in the absence of measurements, that is when $\lambda=0$. 
We choose the initial conditions $\rho_{mn} = \delta_{m,n_0} \delta_{n,n_0}$ and measure the expectation
\be \label{eq:r2} R^2 =\left[ \sum_n (n-n_0)^2 \rho_{nn}\right]/\left[\sum_n \rho_{nn}\right]
\ee
as a function of time. In our evaluation, $n$ ranges from $1$ to $100$, with periodic boundary conditions. $n_0$ is chosen to be $50$. Fig.~\rf{Fig1} shows the result of this evaluation for $\lambda=0$ (a) and $\lambda=1$ (b), for time $t$ ranging from $0$ to $50$. (a) clearly shows localization, while (b) shows diffusion. 
\begin{figure}[tb]
	\includegraphics[width=0.5\textwidth]{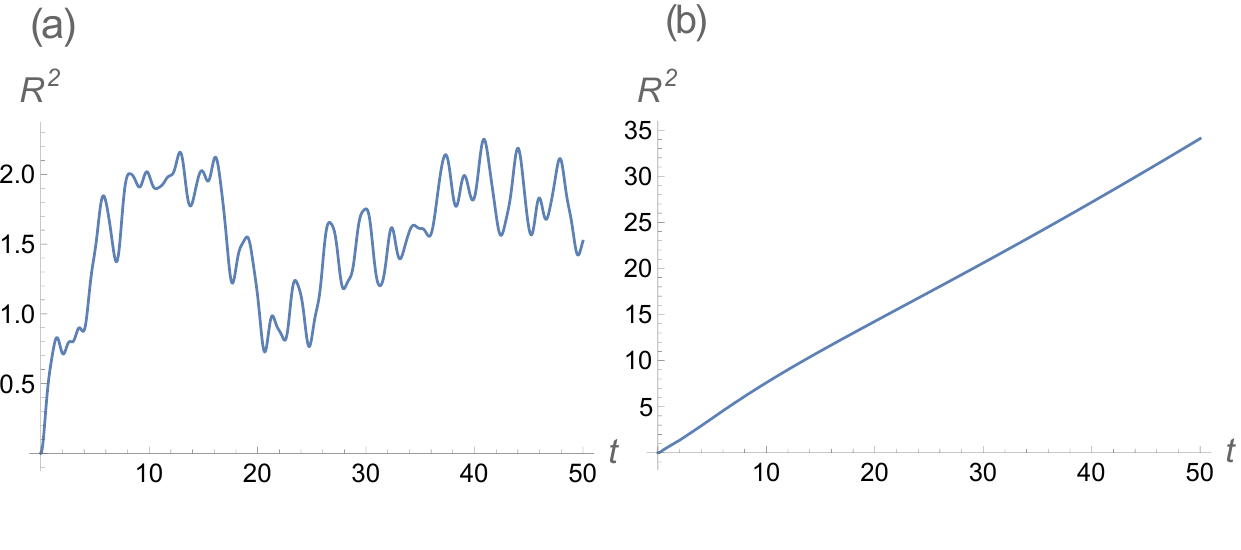}
  \caption{ $R^2$ defined in \rf{eq:r2} for $\lambda=0$ (a) and $\lambda=1$ (b).}
\label{Fig1}
\end{figure}

In conclusion, we developed formalism for studying randomly measured quantum particles which borrow heavily from the theory of time-dependent disorder. It allows to set up this problem in the form of a Schr\"odinger-like equation which in many cases can be solved analytically. Many-particle systems can also be addressed in the form of a suitably constructed Keldysh functional integral \cite{Kamenev2011} with
time dependent disorder. The formalism can also be adapted for computation of non-linear observables such as entanglement entropy by the standard technique of replicating the functional integral
and taking the limit of the number of replicas going to 1. 

The author is grateful to L. Radzihovsky for many comments while this work was being completed, and to L. Pollet for advice concerning the numerical study of the Anderson localized particle. This work was supported by
 the Simons Collaboration on Ultra-Quantum Matter,
which is a grant from the Simons Foundation (651440).



\bibliography{library.bib}

\end{document}